# Incidence Angle and Polarization Dependence of Photo-Induced FMR in Co/Pd Multilayers

J. Saeki, K. Nishibayashi, T. Matsuda, Y. Kitamoto*, and H. Munekata

*Imaging Science and Engineering Laboratory, Tokyo Institute of Technology*
*4259-J3-15 Nagatsuta, Midori-ku, Yokohama 226-8503, Japan*
*\*Interdisciplinary Graduate School of Science and Engineering, Tokyo Institute of Technology*
*4259-J2-40 Nagatsuta, Midori-ku, Yokohama 226-8503, Japan*
E-mail: nishib@isl.titech.ac.jp

Dependence of photo-induced FMR (phi-FMR) on incident angle of excitation and probing laser beams has been studied in a [Co ($d_{Co}$ = 0.78 nm) / Pd ($d_{Pd}$ = 0.81 nm) ]$_5$ multi-layer film with the aim to find experimentally the limitation of inducement and detection of magnetization dynamics with oblique light incidence. We have found, in the experiments changing the incident angle of a pump beam, that phi-FMR is observed up to the grazing incident angle of 88° with *p*-polarized excitation pulses, whereas it disappears at the incidence angle of around 65° with *s*-polarized excitation. As for the experiments changing the incident angle of a probe beam, phi-FMR disappears at the incidence angle of 65° for both *s*- and *p*-polarizations, whereas it reappears with further increasing the angle for the *p*-polarization and vanishes at 75°.

## 1. Introduction

We have proposed the concept of optical signal delay and optical buffer memory on the basis of non-equilibrium state of magnetization caused by the pulsed, optical excitation[1-3]. Preferred device structures which suit for those applications would be a hybrid structure consisting of an optical waveguide and a magnetic thin layer with sufficient magneto-optical (MO) coupling between the two constituent components. In this kind of structure, the excitation of the magnetic layer as well as the detection of excited, *non-equilibrium* magnetization will be carried out via the waveguide/magnet interface by the light which propagates through the optical waveguide. It has been proven through the study of a waveguide-type isolator[4-6] that polarization of light passing through an optical waveguide can be affected significantly by the adjacent magnetic layer. However, the problem of excitation of a magnetic layer with light in a waveguide has never been addressed because device concept which gives motivation to study this fundamental problem is undeveloped. This work aims at studying the excitation of a magnetic layer with a light beam of oblique/grazing incidence, as a simple analogue to the excitation of a magnetic layer with light in a waveguide. For photonic excitation with grazing incidence, energy density of excited area will be influenced significantly by the polarization-dependent reflectivity, referring at least to the Fresnel's law. The shape of the excited area becomes anisotropic as well, which may influence the microscopic process of energy flow between electron and spin subsystems during the inducement of non-equilibrium magnetization as well as propagation of spin waves[7].

With those points in mind, we have investigated for the first time phi-FMR in the ultra-thin Co/Pd multilayers as a function of incident angle of both excitation and detection light pulses for wide range of angle near the grazing condition. This system is known for interface induced perpendicular magnetic anisotropy[8] accompanied by a large magneto-optical Kerr effect[9]. We report in this paper that, in the experiments changing the incident angle of a *pump* beam, phi-FMR can be induced up to the grazing incident angle of 88° with *p*-polarized excitation pulses, whereas it diminishes at the incidence angle of around 65° with *s*-polarized excitation. These behaviors are discussed in terms of the difference in reflectivity in oblique incident angle added with enhanced excitation efficiency found for the *p*-polarized pump beam. As for the experiments changing the incident angle of a *probe* beam, the phi-FMR vanishes at the incidence angle of 65° for both *s*- and *p*-polarizations, whereas it appears again for the *p*-polarization with further increasing the angle but disappears at 75°. Observations in the latter experiment are discussed on the basis of phenomenological electromagnetic picture, using static reflectance and polar Kerr rotation data.

## 2. Experiments

There have been many studies on photo-induced magnetization dynamics in this system with fluence in the region of around mJ/cm$^2$ [10-15]. We studied phi-FMR in ultra-thin [Co/Pd]$_5$ multi-layered structures ($t_{Co}$ = 0.32 ~ 1.03 nm, with $t_{Pd}$ = 0.81 and 1.62 nm) at the region of pump fluences $F$ = 0.11 to 11 $\mu$J/cm$^2$, and reported that the precession amplitude of phi-FMR has increased significantly in a series of samples with $t_{Pd}$ = 0.81 nm, being fifty times larger compared to those in the samples with $t_{Pd}$ = 1.62 nm[16]. Taking into account of such light-sensitive feature, the sample consisting of [Co

($d_{Co}$ = 0.78 nm) / Pd ($d_{Pd}$ = 0.81 nm) ]$_5$ multi-layers was used for the present study. It was prepared by DC magnetron sputtering on the Pd (4.86-nm) / Ta (2.18nm) binary seed layer deposited on a Si (110) substrate at 150 °C [16]. The sample exhibits perpendicular magnetic anisotropy, with coercive force and saturation magnetization field of 120 Oe and 5400 Oe, respectively. The latter value, which was extracted from the value at which the magnetization was saturated along the hard axis[16], is a sum of anisotropy field $H_{ani}$ and the demagnetizing field $H_{dem}$.

Experiment of phi-FMR was carried out by time-resolved magneto-optical (TRMO) spectroscopy on the basis of pump-and-probe technique using a mode-locked Ti: sapphire laser as a light source with wavelength, pulse duration, and repetition of $\lambda$ = 790 nm, $\Delta$ = 90 fs, and $\Gamma$ = 80 MHz, respectively. The fluences of linearly-polarized pump and probe beams were fixed at 7.96 and 0.040 µJ/cm$^2$ per pulse, respectively. The beam size was 200 µm in diameter for both beams when they were focused on the sample surface with normal incidence. The experimental set up of the angle dependent pump-and-probe experiments is shown schematically in Fig. 1. The incident angle of a pump beam, $\theta_{pump}$, was varied between 15 and 89° while keeping the incident angle of a probe beam, $\theta_{probe}$, at $\theta_{probe}$ = 5° for the experiment studying the $\theta_{pump}$ dependence on TRMO temporal profile (Experiment 1). On the other hand, $\theta_{pump}$ was fixed at $\theta_{pump}$ = 2° and the $\theta_{probe}$ value was varied between 26° and 80° for the experiment studying the $\theta_{probe}$ dependence of TRMO temporal profile (Experiment 2). The external magnetic field of $H_{ext}$ = 2000 Oe was applied with the angle of 65°. As stated in ref.16, experiments with this field angle have allowed us to observe and compare systematically the data for both in-plane and out-of-plane samples. Rotation angle of the linearly polarized probe beam was detected by the optical bridge detection apparatus. The detection limit in the present set up was 2.2 µdeg. All the measurements were carried out at room temperature.

The incident-angle dependence of the static polar Kerr rotation was also measured. The difference in the output signals of the calibrated optical bridge between up and down remnant magnetization states was measured as a function of the incident angle of a $s$- or $p$-polarized light beam. The sample was magnetized either up or down by applying vertically the external magnetic field of $H$ = ± 2000 Oe before each optical measurement. The light source used for this measurement was a cw-semiconductor laser with wavelength, power, and beam diameter of $\lambda$ = 785 nm, $P$ = 10 mW, and $d$ = 150 µm, respectively.

### 3. Results and discussion

Shown in the upper two panels of Fig. 2 are temporal TRMO profiles obtained from the Experiment 1 for various pump-pulse incident angle with $s$-polarized (Fig. 2 (a)) and $p$-polarized (Fig.2 (b)) pump beams. For the $s$-polarization, the amplitude of oscillatory MO signals decreases monotonically with increasing $\theta_{pump}$, and has reached to lower detection limit at around $\theta_{pump}$ = 65°. By the way, we are assured that the amplitude of the observed oscillation is proportional to the precession amplitude in the real space at least in the regime of weak excitation[17]. For the $p$-polarization, the amplitude first increases with increasing $\theta_{pump}$ from 15° up to around 55°, beyond which it turns to a gradual reduction. Consequently, the precession has been observable up to $\theta_{pump}$ = 88°. At $\theta_{pump}$ = 89°, the phi-FMR signal has been hardly observed. We model the oscillatory component of phi-FMR with eq.(1), and the amplitude of oscillation, $A_0$, is plotted as a function of $\theta_{pump}$ for both polarization in Fig. 2 (c).

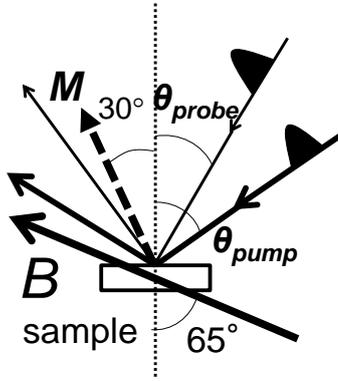

Experiment 1 : $\theta_{pump}$ = 15 - 89° and $\theta_{probe}$ = 5°
Experiment 2 : $\theta_{pump}$ = 2° and $\theta_{probe}$ = 26 - 80°

**Fig. 1** Schematic illustration of experimental set up for angle dependent pump and probe measurement. The sample was mounted on a stationary sample holder, whereas the incident angles of laser beams were varied by adjusting mirrors in optical paths. The ranges of incident angle are depicted in the bottom part of the figure for the Experiment 1 and 2. An external magnetic field of $H_{ext}$ = 2000 Oe was applied with the angle of 65°, under which the direction of magnetization vector was supposed to incline 30° from the axis normal[11].

$$A = A_0 \exp\left(-\frac{t}{\tau}\right) \sin(2\pi f t + \phi) \quad (1)$$

Here, $A_0$, $\tau$, $f$, and $\phi$ are the amplitude of oscillatory MO signals, precession lifetime, precession frequency, and initial phase of the oscillation, respectively. For the experiment with $p$-polarization, the maximum $A_0$ value of 230 µdeg has been obtained at $\theta_{pump}$ = 55°, whereas, at $\theta_{pump}$ = 88°, the value of $A_0$ is still around 30 µdeg which is encouraging in view of optical excitation of a magnetic system with light with grazing incidence or propagating in a waveguide. Besides of $A_0$ value, the precession frequency, precession lifetime, and initial phase are, respectively, $f$ = 8 GHz, $\tau$ = 170 psec, and $\phi$ = −π/2, all of which are not dependent on the $\theta_{pump}$ value.

Regardless of microscopic mechanism of phi-FMR in [Co/Pd] multi-layers, the light power absorbed in the multi-layers per unit volume is expected to vary with $\theta_{pump}$ due to the changes in the reflectivity and the area of excitation. As for the latter, the shape of the area $S$ changes from circle to elliptic with increasing the $\theta_{pump}$ value, being $S = \pi \cdot a \cdot b = \pi \cdot a \cdot a \cdot (\cos \theta_{pump})^{-1}$ where $a$ = 100 μm, the radius of the pump beam impinging normal to the surface.

Shown in the inset Fig. 3 is measured reflectance data as a function of incident angle $\theta_{light}$ for the same [Co/Pd] multi-layer sample. The light source used for the measurement was a cw-semiconductor laser with wavelength, power, and beam diameter of $\lambda$ = 785 nm, $P$ = 30 mW, and $d$ = 150 μm, respectively. The angle $\theta_{light}$ was varied between 3 and 88° for both $s$- and $p$-polarizations. In the data with $p$-polarization, the reflectance minimum appears at $\theta_{light}$ = 78°, indicating the Brewster's angle. The observed steep drop at the angle region larger than 84° for both polarizations is due to the elongation of the light spot along the plane of the incidence, whose length exceeds that of the sample size (~ 4 mm) and results in the reduction in the intensity of the reflected light. Dashed lines in the inset are the fit to measured data on the basis of Fresnel equations. Good fits are obtained with the optical index of $n$ = 2.75 and $\kappa$ = 3.40, which are not abnormal values compared to those of the constituent elements: at $\lambda$ = 790 nm, $n^*(Co) = 3.578 + 4.687\,i$, $n^*(Pd) = 1.998 + 5.048\,i$, $n^*(Ta)$ = 1.130 + 3.448 $i$, and $n^*(Si) = 3.686 + 0.006\,i$.[20]

In order to analyze carefully the experimental data, we introduce eqs.(2a, b) to estimate both the absorbed light power $E_{abs}$ and the efficiency of excitation $\eta$, assuming that (i) the incident light is all absorbed in the 14-nm thick metallic multilayer, and (ii) precession amplitude, and thus $A_0$ value, is proportional to the amount of ultrafast heating by pumping, at least in the regime of weak excitation[16,17].

$$E_{abs}(\theta_{pump}) = F \cdot (1-R) \cdot \cos \theta_{pump} \qquad (2a)$$

$$\eta(\theta_{pump}) = \frac{A_0(\theta_{pump})}{E_{abs}(\theta_{pump})} \Big/ \frac{A_0(0)}{E_{abs}(0)} \qquad (2b)$$

Here, the precession amplitude $A_0(\theta_{pump})$ is given by the value disclosed in Fig.2, the pump fluence $F$ = 7.96 μJ/cm$^2$, and the reflectance $R$ by the calculated curves obtained by fitting the experimental data shown in the inset Fig.3. The values at $\theta_{pump}$ = 0°, and $A_0(0)$, can be safely substituted for values obtained at $\theta_{pump}$ = 15°, since reflectance is nearly constant in the small $\theta_{pump}$ region. The $\eta$ values thus obtained from eqs.(2a, b) are plotted in Fig. 3 as a function of $\theta_{pump}$. For $s$-polarization, the $\eta$ value gradually increases with increasing $\theta_{pump}$ and becomes nearly twice at around $\theta_{pump}$ = 55° compared with that at $\theta_{pump}$ = 15°, beyond which it turns to a reduction. At $\theta_{pump}$ = 65°, at which the amplitude of phi-FMR is barely observable, $\eta \sim 1$ with $E_{abs}$ = 0.7 μJ/cm$^2$. This $E_{abs}$ value is close to the lower bound of phi-FMR with normal pump-and-probe incidence[18]. For $p$-polarization, the $\eta$ value increases with increasing $\theta_{pump}$ up to 88° expect for a small dip at around $\theta_{pump}$ = 60°. At $\theta_{pump}$ = 80°, $\eta \sim 2.5$ using $E_{abs}$ = 1.1 μJ/cm$^2$, and, at $\theta_{pump}$ = 88°, $\eta \sim 4.6$ using $E_{abs}$ = 0.09 μJ/cm$^2$. Here, the values of $E_{abs}$ are those obtained from eq.(2).

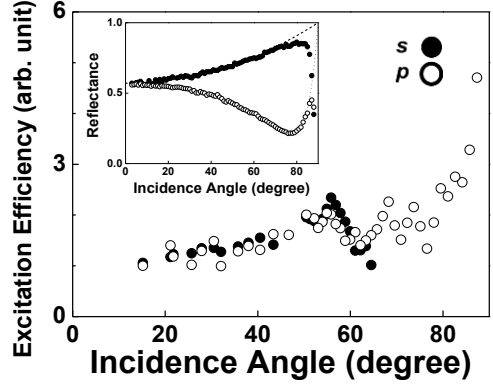

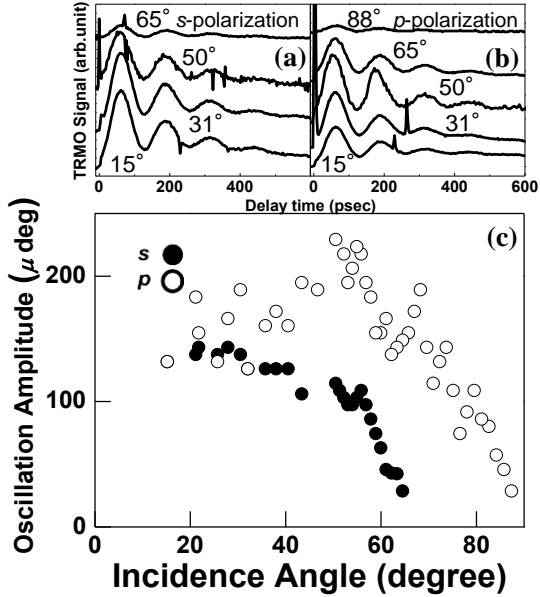

Fig. 2 Temporal profiles of TRMO signals obtained with various $\theta_{pump}$ in the Experiment 1 for (a) $s$-polarization and (b) $p$-polarization. (c) Plots of precession amplitude $A_0$ vs. incident angle $\theta_{pump}$ obtained from Experiment 1 with $s$-polarization (closed-circle) and $p$-polarization (opened-circle). The values of amplitude are extracted by fitting the profiles shown in (a) and (b) with equation (1) mentioned in the main text.

Fig. 3 Excitation efficiency $\eta$ vs. incident angle $\theta_{pump}$ obtained from Experiment 1 with $s$-polarization (closed-circle) and $p$-polarization (opened-circle). for both polarization. Inset shows raw reflectance data obtained for the incident light with $s$-polarization (closed-circle) and $p$-polarization (opened-circle). Dashed lines in the inset are those obtained by fitting the raw data with Fresnel equations.

Although we admit that the calculated value of $R$, and thus $\eta$ and $E_{abs}$, changes steeply in the high $\theta_{pump}$ region, we note here that the $\eta$ value is enhanced in the grazing incidence excitation condition for the $p$-polarization. Referring to the present knowledge as to the phi-FMR in magnetic metals[17,19-21], this means that the efficiency of ultra-fast spin heating (demagnetization)[19] is enhanced when the area of excitation is elongated. We consider here, to account for this hypothetically new effect, a microscopic process based on photo-excited hot carriers ($n^*$) since they are most likely responsible for the ultra-fast demagnetization through spin-flip scattering[22]. For the elongated, excited area, a flow of hot carriers, which is proportional to the gradient $\partial n^*/\partial i$ ($i = x, y$), is reduced along the long ($x$) axis. This will increase the $n^*$ value in the central part of excited area and result in the enhancement in the magnitude of demagnetization in 1 ps.

Results obtained by Experiment 1 indicate that $p$-polarized laser pulses can excite phi-FMR more efficiently than $s$-polarized pulses especially at large incident angle, which is due in part to a large reduction in reflectivity and enhanced $\eta$ value. Increasing the incident angle of pump pulses, however, reduces the areal energy density and thus the amount of temperature increase in the excited area. The competition between those two effects gives rise to the maximum in oscillation amplitude at $\theta_{pump} = 55°$.

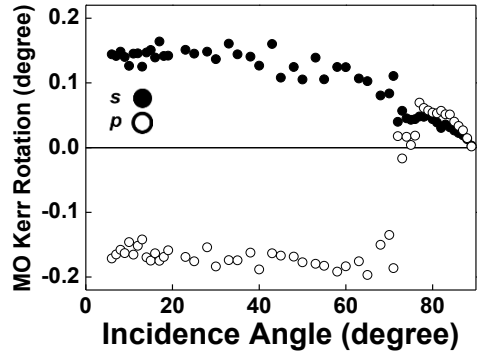

Fig. 5 The magnitude of static polar Kerr rotation as a function of incident angle of a linearly polarized light beam for $s$-polarization (closed-circle) and $p$-polarization (opened-circle).

Let us now show experimental data for the Experiment 2 in Fig. 4. For $s$-polarization, the precession amplitude decreases gradually with increasing $\theta_{probe}$ up to $\theta_{probe} = 55°$, and shows abrupt reduction at $\theta_{probe} = 65°$, beyond which no oscillation has been observed. A spiky temporal MO profile appears at the early stage of time delay instead of the oscillatory TRMO signals. Similar behavior has also been observed in the experiment with $p$-polarized probe. Since magnetization dynamics is not affected by the weak probe pulses, the observed behavior should be attributed to the angle dependent change in the polar Kerr rotation. Shown in Fig. 5 is the magnitude of polar Kerr rotation as a function of incident angle of a linearly polarized light beam. Rotation angle decreases gradually with increases the $\theta_{probe}$ value for $s$-polarization. On the other hand, as for the $p$-polarization, rotation angle is almost unchanged up to around $\theta_{probe} = 70°$, beyond which the sign of rotation switches abruptly from negative to positive, while the magnitude of rotation keeps decreasing with increasing $\theta_{probe}$. These behaviors, which were similar to those reported by early workers[23], can be understood in terms of polarization switching associated with the Brewster's angle. Results obtained by Experiment 2 suggest that the detection angle is restricted to some extent as far as a device utilizing the light in a free space is concerned. Investigation of MO detection in [Co/Pd]- waveguide hybrid structure is desired.

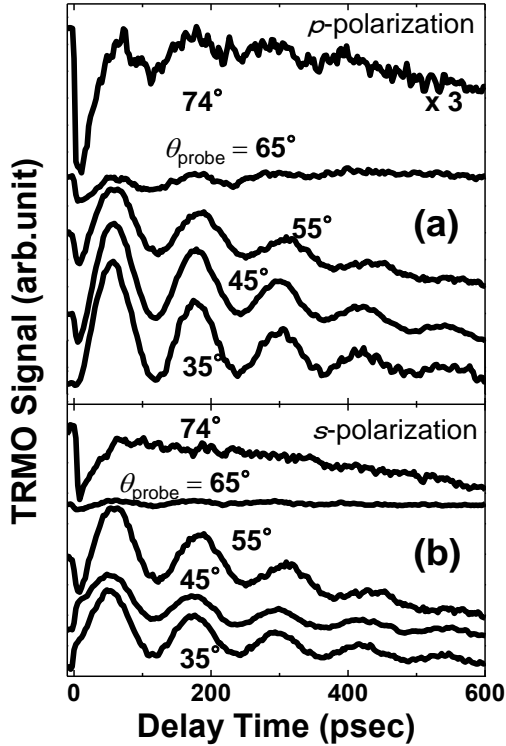

Fig. 4 Temporal profiles of TRMO signals obtained with various $\theta_{probe}$ in the Experiment 2 for (a) $p$-polarization and (b) $s$-polarization. The profile obtained with $\theta_{probe} = 74°$ with $p$-polarization shown in the panel (a) is magnified by a factor of three.

### 4. Conclusions

We have studied dependence of photo-induced FMR in a [Co ($d_{Co} = 0.78$ nm) / Pd ($d_{Pd} = 0.81$ nm)]$_5$ multi-layer film on incident angles of excitation and probing laser beams, with the aim to find experimentally the limitation of inducement and detection of magnetization dynamics with oblique light incidence. The experiment changing the incident angle of a pump beam has revealed that phi-FMR can be induced up to the incident angle of 88° for *p*-polarized excitation, suggesting the enhanced excitation efficiency at grazing incident condition. In contrast, it diminishes at the incidence angle of around 65° with *s*-polarized excitation. These behaviors are discussed in terms of the difference in reflectivity in oblique incidence condition added with enhanced excitation efficiency found for the *p*-polarized pump beam. As for the experiments changing the incident angle of a probe beam, phi-FMR vanishes at the incidence angle of 65° for both *s*- and *p*-polarizations, whereas it reappears for the *p*-polarization with further increasing the angle. At 75°, it finally disappears. The observation in the latter experiment has been explained in terms of the phenomenological electromagnetic wave propagation picture.

**Acknowledgements**  We acknowledge partial supports from Advanced Photon Science Alliance Project from MEXT and Grant-in-Aid for Scientific Research (No. 22226002) from JSPS.

## References

1) H. Munekata, *ICAUMS 2012 and the 36th Annual Congress on MSJ*, Nara, Japan, Oct. 4th (2012).
2) H. Munekata, *MORIS-2009*, Awaji Island, Japan, June 17, (2009).
3) K. Nishibayashi, H. Yoneda, A. Laosunthara, K. Suda, K. Kuga, Y. Hashimoto, and H. Munekata, *J. Mag. Soc. Jpn.* **36**, 74 (2012).
4) M. C. Debnath, V. Zayets, and K. Ando, *Appl. Phys. Lett.* **91**, 043502 (2007).
5) Y. Shoji, T. Mizumoto, H. Yokoi, I-W. Hsieh, and R. M Osgood Jr., *Appl. Phys. Lett.* **92**, 071117 (2008).
6) T. Amemiya, Y. Ogawa, H. Shimizu, H. Munekata and Y. Nakano, *APEX* **1**, 022002 (2008).
7) T. Sato, Y. Terui, R. Moriya, B. A. Ivanov, K Ando, E. Saitoh, T. Shimura, and K. Kuroda, *Nat. Photon.* **6**, 662 (2012).
8) P. F. Carcia, A. D. Meinhaldt, and A. Suna, *Appl. Phys. Lett.* **47**, 178, (1985).
9) Y. Ochiai, S. Hashimoto, and K. Aso, *IEEE Trans. Mag.* **25**, 3755 (1989).
10) A. Barman, S. Wang, O. Hellwig, A. Berger, E. E. Fullerton, and H. Schmidt, *J. Appl. Phys.* **101**, 09D102 (2007).
11) S. Mizukami, E. P. Sajitha, D. Watanabe, F. Wu, T. Miyazaki, H. Naganuma, M. Oogane, and Y. Ando, *Appl. Phys. Lett.*, **96**, 152502 (2010).
12) E. P. Sajitha, J. Walowski, D. Watanabe, S. Mizukami, F. Wu, H. Naganuma, M. Oogane, Y. Ando, and T. Miyazaki, *IEEE. Trans. Mag.* **46**, 2056 (2010).
13) Z. Liu, R. Brandt, O. Hellwig, S. Florez, T. Thomson, B. Terris, and H. Schmidt, *J. Magn. Magn. Mater.* **323**, 1623 (2011).
14) S. Pal, B. Rana, O. Hellwig, T. Thomson, and A. Barman, *Appl. Phys. Lett.* **98**, 082501 (2011).
15) T. Kampfrath, R. G. Ulbrich, *Phys. Rev. B* **65**, 104429 (2002)
16) K. Yamamoto, T. Matsuda, K. Nishibayashi, Y. Kitamoto and H. Munekta, *IEEE Trans Mag.* **49**, 3155 (2013)
17) K. Nishibayashi, K. Kuga, and H. Munekata, *AIP Adv.* **3**, 032107 (2013).
18) http://www.sspectra.com/sopra.html (cited on Dec. 20, 2013.)
19) E. Beaurepaire, J.-C. Merle, A. Daunois, J.-Y. Bigot, *Phys. Rev. Lett.* **76**, 4250 (1996).
20) B. Koopmans, "Laser-induced magnetization dynamics", *Spin Dynamics in Confined Magnetic Structures II*, B. Hillebrands and K. Ounadjela, Eds., Springer, 2003, pp. 253-316.
21) A. Kirilyuk, A. V. Kimel, T. Rasing, *Rev. Mod. Phys.* **82,** 2731 (2010).
22) M. Cinchetti, M. S. Albaneda, D. Hoffmann, T. Roth, J. –P.Wüstenberg, m. Krauβ, O. Andreyev, H. C. Schneider, M. Bauer, and M. Aeschlimann, *Phys. Rev. Lett.* **97**, 177201 (2006).
23) C. Y. You and S. C. Shin, *Appl. Phys. Lett.* **69**, 1315 (1996).